\documentclass[twocolumn,preprintnumbers,showpacs,superscriptaddress]{revtex4}

\usepackage{amsmath}
\usepackage[pdftex]{graphicx}
\usepackage{dcolumn}
\usepackage{bm}
\usepackage[latin1]{inputenc}
\usepackage{color}
\usepackage{SIunits}

\newcommand{\ud}[1]{{#1^{\dagger}}}

\begin{document}

\title{Exciton-polariton mediated superconductivity}
\author{Fabrice P.~Laussy}
\affiliation{School of Physics and Astronomy, University of Southampton, Southampton, SO171BJ, United Kingtom}
\author{Alexey V. Kavokin}
\affiliation{School of Physics and Astronomy, University of Southampton, Southampton, SO171BJ, United Kingtom}
\affiliation{Dipartimento di Fisica, Universita' di Roma ``Tor Vergata'', 00133 Roma, Italy}
\author{Ivan A.~Shelykh}
\affiliation{Science Institute, University of Iceland, Dunhagi 3, IS-107, Reykjavik, Iceland}
\affiliation{St.~Petersburg Academic University, Khlopina str., 8/3, 194021 St.~Petersburg, Russia}
\date{\today }

\begin{abstract}
  We revisit the exciton mechanism of superconductivity in the
  framework of microcavity physics, replacing virtual excitons as a
  binding agent of Cooper pairs by excitations of an exciton-polariton
  Bose-Einstein condensate.  We consider a model microcavity where a
  quantum well with a two dimensional electron gas is sandwiched
  between two undoped quantum wells, where a polariton condensate is
  formed. We show that the critical temperature for superconductivity
  dramatically increases with the condensate population, opening a new
  route towards high temperature superconductivity.
\end{abstract}

\maketitle



\begin{figure}[t]
\centering
\includegraphics[width=\linewidth]{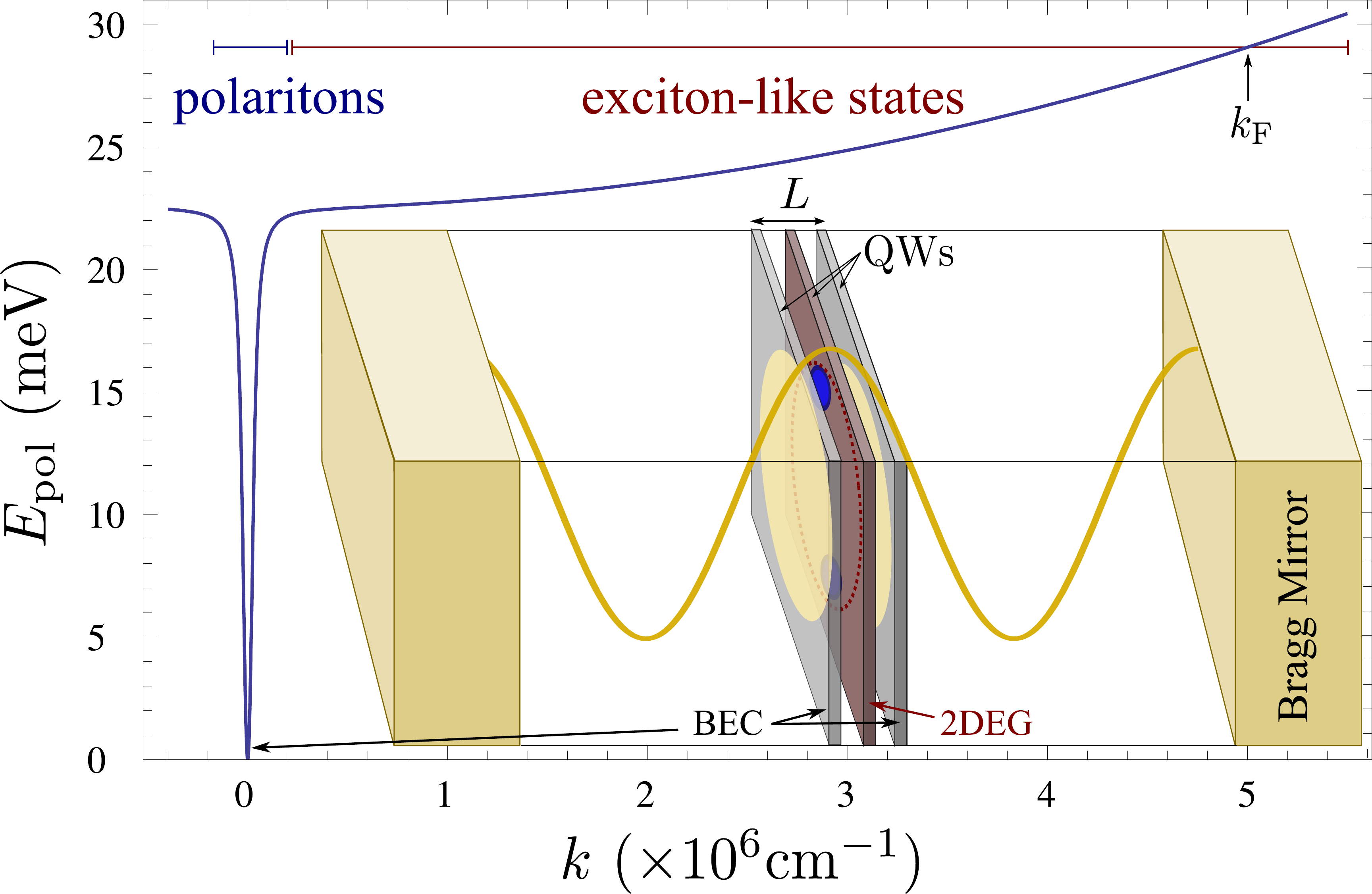}
\caption{(Colour online) The polariton dispersion, on the scale of
  interest for mediating Cooper-pairing at
  $k_\mathrm{F}=\unit{5\times10^6}\centi\meter^{-1}$. The photon
  (electron/hole) effective mass has been taken as $10^{-5}m_0$
  ($0.22/1.25m_0$), $m_0$ being the electron mass in vacuum, the
  vacuum Rabi splitting as~$\Omega =45$meV, $X^2=1/2$ and
  $UA\approx0.6\protect\mu $eV$\times \protect\mu m^{2}$. Other
  parameters are~$L=50Å$, $a_\mathrm{B}=25Å$, $R_y=32$meV,
  $\epsilon\approx7\epsilon_0$. The peculiar polariton dispersion
  allows for the coexistence of a BEC at low~$k$ that mediate the
  interaction and of slow exciton-like states at large~$k$ providing
  the retarded electron-electron attraction. In inset, the scheme of
  the model microcavity structure with an $n$-doped QW sandwiched
  between two undoped QWs where the polariton condensate is formed.}
\label{fig:MonJun22034820BST2009}
\end{figure}

Microcavity polaritons are quasiparticles that arise from the strong
coupling of microcavity photons with quantum-well (QW)
excitons~\cite{kavokin_book07a}. They have attracted considerable
interest for their manifestations of quantum phenomena, from
stimulated scattering~\cite{savvidis00a} and polariton
lasing~\cite{deng02a} to Bose-Einstein condensation
(BEC)~\cite{kasprzak06a} and superfluidity~\cite{amo09a,amo09b}. Their
in-plane dispersion is strongly nonparabolic. Near the ground state
corresponding to zero in-plane wavevector, an extremely small
polariton mass---averaging the exciton mass and the much smaller
cavity-photon mass---brings the critical temperature for quantum
degeneracy up to room temperature~\cite{kavokin03a,baumberg08a}. On
the other hand, at wavevectors exceeding the wavevector of light at
the exciton resonance frequency, the polariton dispersion becomes
exciton-like and its effective mass exceeds that near the band minimum
by four orders of magnitude (see
Fig.~\ref{fig:MonJun22034820BST2009}). Exciton-polaritons are
electrically neutral and cannot carry electric current. However, they
may coexist and interact with free electrons or holes, if these
carriers are introduced in the same QW with the excitons or in a
neighboring QW~\cite{lagoudakis03a}. When confined together,
exciton-polaritons and free carriers form a Bose-Fermi mixture which
is expected to exhibit peculiar optical and electronic properties. In
this Letter, we study the possibility of superconductivity in
semiconductor microcavities containing both undoped QWs and thin
$n$-doped semiconductor layers. We show that exciton-polariton
mediated superconductivity is possible.

Conventional superconductivity occurs at low temperatures and can be
described within the framework of BCS theory~\cite{bardeen57b}, which
relies on the formation of Cooper pairs. Following the discovery of a
phonon-mediated attraction between electrons~\cite{bardeen51a}, Cooper
found that two electrons on top of a Fermi sea always form a bound
state, however small the (attractive) interaction between them~%
\cite{cooper56a}. This instability results in the so-called BCS state
(coherent state of Cooper pairs), that leads to a gap in the spectrum
of excitations, responsible for superconductivity. In the currently
available high-$T_{\mathrm{C}}$ superconductors (the cuprates), an
electron pairing is also thought to be realized through a mediating
boson, although probably not the phonon~\cite{giustino08a}. Excitons
have been proposed as suitable mediating bosons to achieve higher
critical temperatures of superconductivity in specially designed
heterostructures~(\cite{little64a}, see~\cite{ginzburg_book09a} for a
review). As compared to phonons, the characteristic cut-off energy
above which the attractive character of the interaction is lost for
the excitons is several orders of magnitude higher and the critical
temperature is therefore expected to be sufficiently increased with
respect to the BCS superconductors. One possible implementation of
this idea is the so-called sandwich type of superconducting system,
consisting of layers of metal and insulator materials where excitons
and free electrons are close enough to each other to interact
efficiently~\cite{allender73a}. Electrons in metal are attracted to
each other due to double scatterings involving creation and
annihilation of virtual excitons of the semiconductor. The virtual
excitons are created due to the non energy-conserving scattering
process and disappear also due to scattering with free electrons from
the metal. However, the probability of such scattering is low, because
of the huge energies needed to create virtual excitons. While the
metal-insulator sandwiches demonstrated superconductivity at
temperatures up to 50K~\cite{gozar08a,aruta08a}, there is no evidence
of the exciton mechanism of superconductivity being realised in this
or other systems.

Relying on the recent discovery of high-temperature BEC of
exciton-polaritons, we propose microcavities as an implementation of
the Ginsburg scheme of mediating the interaction between electrons
through an exciton-like field ~\cite{ps_ginsburg77}, using not virtual
but real excitons. We show that the strength of electron-electron
interactions, mediated by a condensate of exciton-polaritons, scales
like the density of the polariton condensate $N_{0}$, which can be
tuned by the pumping power. This opens the route to optically mediated
superconductivity, where the BCS gap value is governed by an optical
pumping intensity. We consider a device specially engineered for this
purpose (see the inset of Figure~\ref%
{fig:MonJun22034820BST2009}), consisting of an $n$-doped QW sandwiched
between two undoped QWs at the antinode of the optical field of a
microcavity. A polariton condensate is formed in the sandwiching
wells, e.g., by optical excitation. The advantage of this design is to
maximize the interaction of the 2DEG with the condensate, which, being
delocalized in the structure (thanks to the photonic part of the
polariton), can sandwich as a single object the electron gas. In this
way, proximity of polaritons and electrons is maximum. Also, a larger
number of QWs allows to reach higher condensate densities of
polaritons. In the rest of this Letter, we derive an effective
electron-electron Hamiltonian from the electron-condensate
interaction, show the existence of an effective attractive potential
that grows with the condensate occupancy, and solve numerically the
gap equation for the polariton-mediated superconductivity.

The Hamiltonian of the system reads (denoting fermionic operators
corresponding to electrons of in-plane momentum~$\hbar \mathbf{k}$ by
$%
\sigma _{\mathbf{k,}}$ and bosonic operators corresponding to polaritons by
$a_{\mathbf{k}}$):
\begin{multline}
  H=\sum_{\mathbf{k}}\left[ E_{\mathrm{el}}(\mathbf{k}){\sigma
      _{%
        \mathbf{k}}^{\dagger }}\sigma _{\mathbf{k}
    }+E_{\mathrm{pol}}(%
    \mathbf{k}){a_{\mathbf{k}}^{\dagger }}a_{\mathbf{k}}\right] + \\
  +\sum_{\mathbf{k}_{1},\mathbf{k}_{2},\mathbf{q}}\Bigg[
V_\mathrm{C}(\mathbf{q})\ud{\sigma_{\mathbf{k}_1+\mathbf{q}}}\ud{\sigma_{\mathbf{k}_2-\mathbf{q}}}\sigma_{\mathbf{k}_1}\sigma_{\mathbf{k}_2,}
 \\
  +XV_\mathrm{X}(\mathbf{q}){\sigma
    _{\mathbf{k}_{1}}^{\dagger }}\sigma
  _{\mathbf{k}_{1}+\mathbf{q}%
    }{a_{\mathbf{k}_{2}+\mathbf{q}}^{\dagger }}a_{\mathbf{k}_{2}}
  +U{a_{\mathbf{k}%
      _{1}}^{\dagger }}{a_{\mathbf{k}_{2}+\mathbf{q}}^{\dagger
    }}a_{\mathbf{k}_{1}+%
    \mathbf{q}}a_{\mathbf{k}_{2}}\Bigg]\,.  \label{eq:MonJul13163608BST2009}
\end{multline}

Here $E_{\mathrm{el}}(\mathbf{k})$ and $E_{\mathrm{pol}}(\mathbf{k})$
describe the in-plane dispersion of electrons and exciton-polaritons,
respectively. 
The third term describes the direct electron-electron interaction, the
fourth the electron-polariton interaction and the fifth
polariton-polariton interactions which are treated within the $s$-wave
scattering approximation with strength $U=6a_{\mathrm{B}
}^{2}R_{y}X^{2}/A$ (where $a_{\mathrm{B}}$ is the exciton Bohr radius,
$R_{y}$ the exciton binding energy and $A$ the normalization area, $X
$ is the exciton Hopfield coefficient, which square quantifies the
exciton fraction in the exciton-polariton
condensate)~\cite{tassone99a}. We go beyond BCS theory by including
the electron-electron Coulomb interaction $V_\mathrm{C}$ on top of the
electron-exciton interactions $V_\mathrm{X}$. Their respective matrix
elements are given by $V_\mathrm{C}(\mathbf{q})=e^2/[{2\epsilon
  A}(|\mathbf{q}|+\kappa)]$
and:
\begin{multline}
  \label{eq:WedApr29165133BST2009} 
  V_\mathrm{X}(\mathbf{q})=\frac{16e^{2}}{A\epsilon
    a_{\mathrm{B}}^{3}}\frac{\pi
    e^{-|\mathbf{q}|L/2}}{|\mathbf{q}|+\kappa_\mathbf{q}}\times
\\
\left\{ \frac{1}{\beta_{h}^{2}}\frac{1}{\left( |\mathbf{q}|^{2}+\frac{4}{a_{%
\mathrm{B}}^{2}\beta _{h}^{2}}\right) ^{3/2}}-\frac{1}{\beta _{e}^{2}}\frac{1%
}{\left( |\mathbf{q}|^{2}+\frac{4}{a_{\mathrm{B}}^{2}\beta _{e}^{2}}\right)
^{3/2}}\right\} \,,
\end{multline}
where~$\beta_{e(h)}=m_{e(h)}/(m_{e}+m_{h})$~\cite{ciuti98a}, $\kappa$
is the screening wavelength in-plane for the electrons in the middle
QW, which is estimated as~$\kappa=m_ee^2/(2\pi\epsilon\hbar^2)$ (its
weak temperature dependence can be ignored) with~$n$ the density of
the two-dimensional electron gas (2DEG) and
$\kappa_\mathbf{q}=\kappa\exp(-|\mathbf{q}|L/2)$ is derived from the
Lindhard formula~\cite{haug_book90a} in the geometry of
Fig.~\ref{fig:MonJun22034820BST2009}. Here we neglected all exchange
terms assuming that the overlap of electron and exciton wavefunctions
is zero. We use the mean-field approximation for the condensate of
exciton-polaritons, namely, $a_{\mathbf{k}_{1}+\mathbf{q}%
}^{\dagger }a_{\mathbf{k}_{1}}\approx \langle
a_{\mathbf{k}_{1}+\mathbf{q}%
}^{\dagger }\rangle a_{\mathbf{k}_{1}}+a_{\mathbf{k}_{1}+\mathbf{q}%
}^{\dagger }\langle a_{\mathbf{k}_{1}}\rangle $ and~$\langle
a_{\mathbf{k}%
}\rangle \approx \sqrt{N_{0}A}\delta _{\mathbf{k},\mathbf{0}}$ with
$N_{0}$ the density of polaritons in the condensate. 
This allows us to obtain the following expression for the Hamiltonian,
after diagonalizing the polariton part by means of a Bogoliubov
transformation (that leaves the free propagation of electrons and
their direct interaction, $H_\mathrm{C}$, invariant):
\begin{multline}
  H=\sum_{\mathbf{k}}E_{\mathrm{el}}(\mathbf{k}){\sigma_{\mathbf{k}}^{\dagger }}\sigma _{\mathbf{k}}+\sum_{\mathbf{k}}E_{%
    \mathrm{bog}}(\mathbf{k}){b_{\mathbf{k}}^{\dagger }}b_{\mathbf{k}}+ \\
  +H_\mathrm{C}+\sum_{\mathbf{k},\mathbf{q}}M(\mathbf{q}){\sigma _{\mathbf{k}}^{\dagger }}\sigma _{\mathbf{k}+\mathbf{q}}({b_{-\mathbf{q}%
    }^{\dagger }}+b_{\mathbf{q}})\,  \label{Hamiltonian_3}
\end{multline}%
where $E_{\mathrm{bog}}(\mathbf{k})$ describes the dispersion of the
elementary excitations (bogolons) of the interacting Bose gas, which
is very close to a parabolic exciton dispersion at large~$k$:
\begin{equation}
E_{\mathrm{bog}}(\mathbf{k})=\sqrt{\tilde{E}_{\mathrm{pol}}(\mathbf{k})(%
\tilde{E}_{\mathrm{pol}}(\mathbf{k})+2UN_{0}A)}\,,
\end{equation}%
where $\tilde{E}_{\mathrm{pol}}(\mathbf{k})\equiv E_{\mathrm{pol}}(\mathbf{k}%
)-E_{\mathrm{pol}}(\mathbf{0})$ and with the renormalized bogolon-electron
interaction strength: 
\begin{equation}
  M(\mathbf{q})=\sqrt{N_{0}A}XV_\mathrm{X}(\mathbf{q})\sqrt{\frac{E_{\mathrm{bog}}(\mathbf{%
        q})-\tilde{E}_{\mathrm{pol}}(\mathbf{q})}{2UN_{0}A-E_{\mathrm{bog}}(\mathbf{q%
      })+\tilde{E}_{\mathrm{pol}}(\mathbf{q})}}\,.  \label{Frohlich}
\end{equation}

The last term of Eq.~(\ref{Hamiltonian_3}) coincides with the Frölich
electron-phonon interaction Hamiltonian, which allows us to write an
effective Hamiltonian for the bogolon-mediated electron-electron
interaction. This results in an effective interaction between
electrons, of the type $\sum_{\mathbf{k}_{1},
  \mathbf{k}_{2},\mathbf{q}}V_{\mathrm{eff}}(\mathbf{q},\omega
){\sigma _{ \mathbf{k}_{1}}^{\dagger }}\sigma
_{\mathbf{k}_{1}+\mathbf{q}}{\sigma}_{\mathbf{k}_{2}+\mathbf{q}}^{\dagger}\sigma _{\mathbf{k}_{2}}$, with
$\hbar\omega
=E_{\mathrm{pol}}(\mathbf{k}_{1}+\mathbf{q})-E_{\mathrm{pol}}(%
\mathbf{k}_{1})$ and the effective interaction strength
$V_{\mathrm{eff}}(\mathbf{q},\omega
)=V_\mathrm{C}(\mathbf{q})+V_\mathrm{A}(\mathbf{q},\omega)$, with:
\begin{equation}
  V_\mathrm{A}(\mathbf{q},\omega)=\frac{2M(\mathbf{q})^{2}E_{\mathrm{bog}}(\mathbf{q})}{(\hbar \omega )^{2}-E_{\mathrm{bog}}(\mathbf{q})^{2}}\,.
\label{V_eff}
\end{equation}
\begin{figure}[t]
\centering
\includegraphics[width=\linewidth]{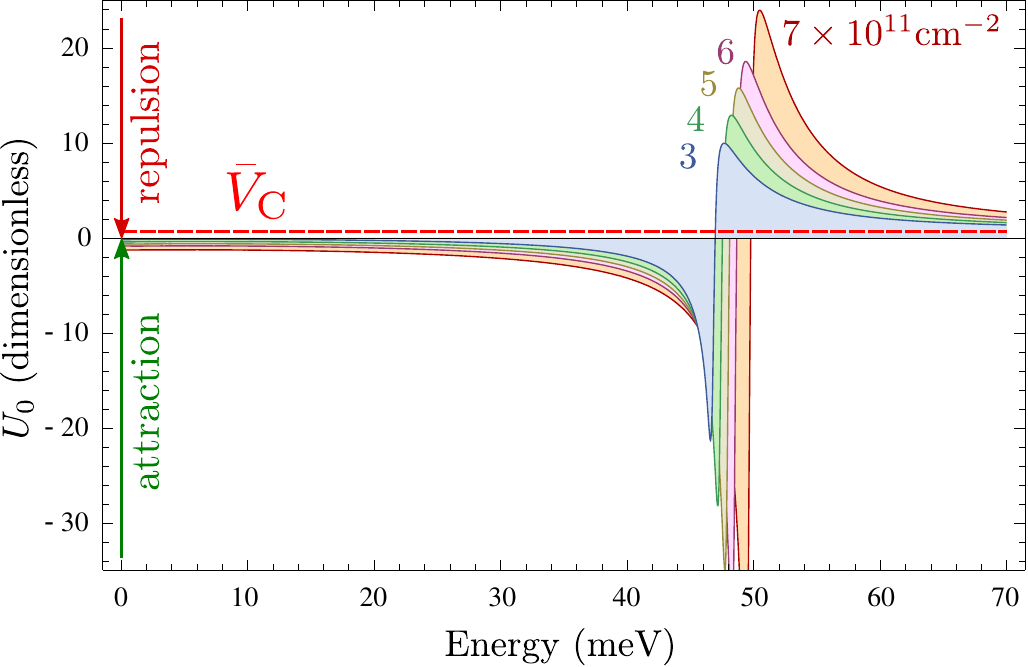}
\caption{(Colour online) The effective electron-electron interaction
  potential $U_0$ (cf.~Eq.~(\protect\ref%
  {eq:ThuApr30131709BST2009})) for different values of the condensate
  density~$N_0$ (in $\centi\meter^{-2}$). Direct Coulomb repulsion is
  included (horizontal red dashed line) which fights the fragile
  retardation effect of excitons. Attraction is restored when the
  density of the condensate is high enough to overcome Coulomb
  repulsion, leading to Cooper pairing in the 2DEG.}
\label{fig:MonJun22034817BST2009}
\end{figure}
Equation~(\ref{V_eff}) recovers the boson-mediated interaction
potential obtained for Bose-Fermi mixture of cold atomic
gases~\cite{bijlsma00a}, in the limit of vanishing exchanged
wavevectors. It describes the BEC\ induced attraction between
electrons. Remarkably, it increases linearly with the condensate
density $N_{0}$. This represents an important advantage of this
mechanism of superconductivity with respect to the earlier proposals
of exciton-mediated
superconductivity~\cite{little64a,ginzburg_book09a,allender73a}, as
the strength of Cooper coupling can be directly controlled by optical
pumping of the exciton-polariton condensate.

The frequency dependence in Eq.~(\ref{V_eff}) describes the retarded
character of the mediated electron-electron attraction, which is
essential for the stability of a Cooper pair. In general, for the
exciton-induced interactions, the retardation is weak as compared to
that provided by the phonon-induced interactions. In our system the
exciton mass is only six times heavier than the electron mass, which
results in a speed of bogolon excitations six times smaller than the
Fermi velocity. The weak retardation is compensated by the strong
polariton-mediated attraction in the case of a high density of the
condensate.

\begin{figure}[t]
\centering
\includegraphics[width=\linewidth]{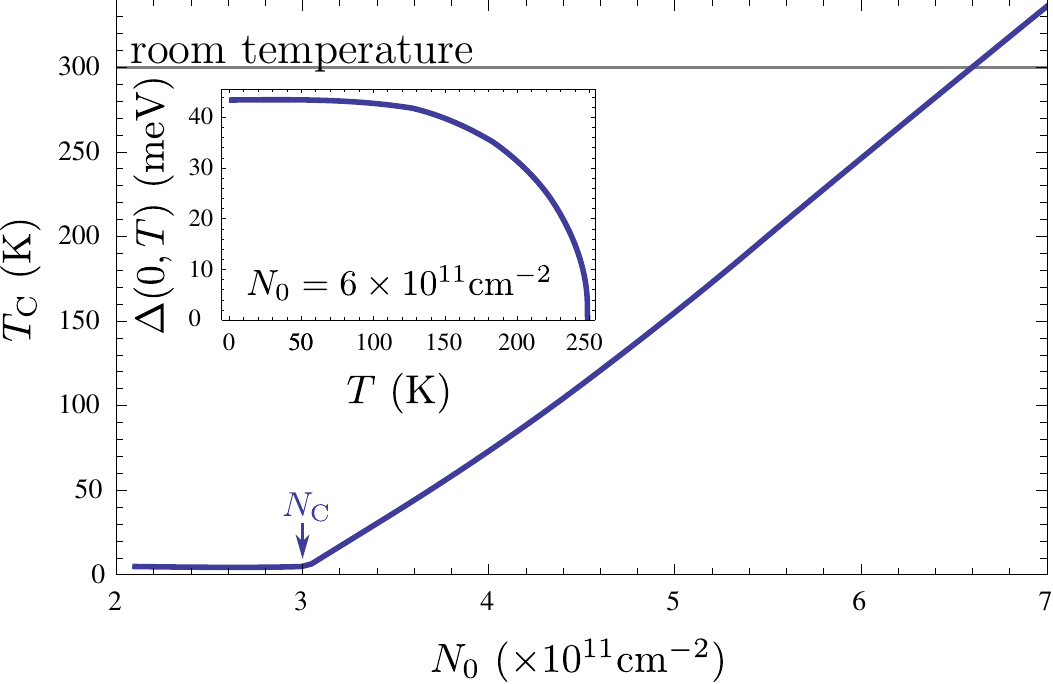}
\caption{Critical temperature of superconductivity as a function of
  the exciton-polariton condensate density $N_0$. Inset: closing of
  the gap at the energy of the Fermi sea for
  $N_0=6\times10^{11}$cm$^{-2}$.}
\label{fig:SatJun20122420BST2009}
\end{figure}

We compute an effective (dimensionless) interaction between electrons
by averaging the potential of interaction~$V_\mathrm{eff}$ over the 2D
Fermi sea, that includes the attractive exciton mechanism
($V_\mathrm{A}$) competing with direct Coulomb repulsion
($V_\mathrm{C}$)~\cite{ginzburg_book09a}:
\begin{equation}
  \label{eq:ThuApr30131709BST2009}
  U_{0}(\omega )=\frac{A\mathcal{N}}{2\pi}\int_{0}^{2\pi}[V_\mathrm{A}(q,\omega)+V_\mathrm{C}(q)]d\theta \,,  
\end{equation}
where~$q=\sqrt{2k_\mathrm{F}^2(1+\cos\theta)}$,
$\mathcal{N}=dn/dE|_{E_{\mathrm{F}}}= m_{e}/(\pi \hbar ^{2})$ is the
density of states at the Fermi surface of the 2DEG. The shapes of
$U_0(\omega)$ for various polariton densities are shown in
Fig.~\ref{fig:MonJun22034817BST2009}. Coulomb repulsion contributes an
overall (frequency independent) repulsion $\bar V_\mathrm{C}$, of
order $0.67$ (it is dimensionless) for our parameters. This repulsion
is detrimental to the exciton mechanism, which has a fragile
retardation effect compared to metals. At the density of
$N_0=\unit{2\times10^{11}}\centi\meter^{-2}$, for instance, the effective
electron-electron potential is dominated by Coulomb repulsion and thus
becomes repulsive again at long times. It exhibits attraction only in
a narrow region of frequencies, that is not sufficient to bind Cooper
pairs.  At higher densities, retardation extends over longer time,
like in the BCS picture.  We compute the average of $V_\mathrm{A}$
numerically and solve the gap equation for~$U_0$
\begin{equation}
  \label{eq:SatJun20114247BST2009}
  \Delta (\xi ,T)=-\int_{-\infty }^{+\infty }\frac{U_{0}(\xi -\xi')\Delta (\xi' ,T)\text{tanh}(E/2k_\mathrm{B}T)}{2E}%
d\xi'
\end{equation}
by iteration, where~$E=\sqrt{\Delta(\xi',T)^2+\xi'^2}$. In inset of
Fig.~\ref{fig:SatJun20122420BST2009}, $\Delta(0,T)$ is displayed for
$N_0=\unit{6\times10^{11}}\centi\meter^{-2}$. A gap is open on the
Fermi sea, that results in the Cooper instability that drives the
ground state to a BCS-like state. 
%
%
The gap is suppressed by temperature, but its value and the critical
temperature $T_\mathrm{C}$, at which the gap vanishes, strongly depend
on $N_0$, as shown in Fig.~\ref{fig:SatJun20122420BST2009}. We have
used the numerical parameters typical of GaN based microcavities (see
caption of Fig.~\ref{fig:MonJun22034820BST2009}), where polariton
lasing at room temperature has been recently
reported~\cite{christmann08a}. The free electron concentration has
been chosen as $n\approx 4\times 10^{12}$cm$^{-2}$, so
that~$k_{\mathrm{F}}=5\times10^{6}$cm$^{-1}$ falls in the exciton part
of the polariton dispersion
(cf.~Fig.~\ref{fig:MonJun22034820BST2009}). The lateral size of the
polariton-mediated Cooper pairs can be estimated as $\chi
_{\mathrm{C}}\approx (2\pi
\hbar^{2}k_{\mathrm{F}})/(m_{e}\Delta(0,0))$.  At the highest power
reported here, $\chi_{\mathrm{C}}\approx 196\mathring{A} $ is well
above the average distance between electrons, $n^{-1/2}\approx
25\mathring{A}$. The parameters typical of the GaAs based
microcavities give essentially similar results, while in GaAs-based
cavities the polariton BEC cannot exist at temperatures higher than
several tens of Kelvin, so that superconducting $T_{\mathrm{C}}$ would
not exceed this value.

The critical temperature for superconductivity is zero for low
exciton-polariton concentrations because Coulomb repulsion of
electrons prevents formation of the Cooper pairs. The
electron-electron interaction is repulsive at low frequencies in this
case. Above a critical density, which is slightly
below~$N_{\mathrm{C}}\approx3\times\unit{10^{11}}\centi\meter^{-2}$ in
our model system, the superconductivity becomes possible.  The
electron-electron interaction strength dependence on the polariton
density leads to a roughly linear increase of the critical temperature
$T_{\mathrm{C}}$ with~$N_{0}$ in this region. This is characteristic
for systems with strong electron-electron interactions. Tuning the
condensate density with external power allows a transition to the
superconducting regime, up to very high temperatures (limited by the
Mott transition for the exciton-polariton condensate expected at about
$N_0=10^{12}$cm$^{-2}$ in GaN based microcavities~\cite{kavokin03a});
with a GaN structure, this would enable superconductivity up to room
temperature.

In order to evidence experimentally the light-mediated
superconductivity, one could measure the in-plane differential
photoconductivity of the microcavity. The carriers need to be injected
in the $n$-doped quantum well from metallic contacts. The
polariton condensate may be created by resonant optical pumping. The
sign of differential photoconductivity is expected to change from
negative to positive at the onset of superconductivity.

In conclusion, we propose a new mechanism to achieve
superconductivity, based on microcavity polaritons. A Bose-Einstein
condensate of polaritons is offered as a mediator for the interactions
between electrons in a specially engineered device. The magnitude of
attraction increases linearly with the condensate density, allowing
for an external control of the binding energy, and therefore of the
critical temperature. With devices that have demonstrated polariton
BEC up to room temperature, our findings suggest that
exciton-polaritons could be promising candidates to achieve
high-temperature superconductivity in a semiconductor structure, with
critical temperatures only limited by that of the BEC.

We thank T. Taylor for help in numerical modeling. FPL and AVK
acknowledge support from the EPSRC and AVK from EU e-I3 ETSF project
211956.

\bibliography{books,Sci,pol-sup}

\begin{thebibliography}{24}
\expandafter\ifx\csname natexlab\endcsname\relax\def\natexlab#1{#1}\fi
\expandafter\ifx\csname bibnamefont\endcsname\relax
  \def\bibnamefont#1{#1}\fi
\expandafter\ifx\csname bibfnamefont\endcsname\relax
  \def\bibfnamefont#1{#1}\fi
\expandafter\ifx\csname citenamefont\endcsname\relax
  \def\citenamefont#1{#1}\fi
\expandafter\ifx\csname url\endcsname\relax
  \def\url#1{\texttt{#1}}\fi
\expandafter\ifx\csname urlprefix\endcsname\relax\def\urlprefix{URL }\fi
\providecommand{\bibinfo}[2]{#2}
\providecommand{\eprint}[2][]{\url{#2}}

\bibitem[{\citenamefont{Kavokin et~al.}(2007)\citenamefont{Kavokin, Baumberg,
  Malpuech, and Laussy}}]{kavokin_book07a}
\bibinfo{author}{\bibfnamefont{A.}~\bibnamefont{Kavokin}},
  \bibinfo{author}{\bibfnamefont{J.~J.} \bibnamefont{Baumberg}},
  \bibinfo{author}{\bibfnamefont{G.}~\bibnamefont{Malpuech}}, \bibnamefont{and}
  \bibinfo{author}{\bibfnamefont{F.~P.} \bibnamefont{Laussy}},
  \emph{\bibinfo{title}{Microcavities}} (\bibinfo{publisher}{Oxford University
  Press}, \bibinfo{year}{2007}).

\bibitem[{\citenamefont{Savvidis et~al.}(2000)\citenamefont{Savvidis, Baumberg,
  Stevenson, Skolnick, Whittaker, and Roberts}}]{savvidis00a}
\bibinfo{author}{\bibfnamefont{P.~G.} \bibnamefont{Savvidis}},
  \bibinfo{author}{\bibfnamefont{J.~J.} \bibnamefont{Baumberg}},
  \bibinfo{author}{\bibfnamefont{R.~M.} \bibnamefont{Stevenson}},
  \bibinfo{author}{\bibfnamefont{M.~S.} \bibnamefont{Skolnick}},
  \bibinfo{author}{\bibfnamefont{D.~M.} \bibnamefont{Whittaker}},
  \bibnamefont{and} \bibinfo{author}{\bibfnamefont{J.~S.}
  \bibnamefont{Roberts}}, \bibinfo{journal}{Phys. Rev. Lett.}
  \textbf{\bibinfo{volume}{84}}, \bibinfo{pages}{1547} (\bibinfo{year}{2000}).

\bibitem[{\citenamefont{Deng et~al.}(2002)\citenamefont{Deng, Weihs, Santori,
  Bloch, and Yamamoto}}]{deng02a}
\bibinfo{author}{\bibfnamefont{H.}~\bibnamefont{Deng}},
  \bibinfo{author}{\bibfnamefont{G.}~\bibnamefont{Weihs}},
  \bibinfo{author}{\bibfnamefont{C.}~\bibnamefont{Santori}},
  \bibinfo{author}{\bibfnamefont{J.}~\bibnamefont{Bloch}}, \bibnamefont{and}
  \bibinfo{author}{\bibfnamefont{Y.}~\bibnamefont{Yamamoto}},
  \bibinfo{journal}{Science} \textbf{\bibinfo{volume}{298}},
  \bibinfo{pages}{199} (\bibinfo{year}{2002}).

\bibitem[{\citenamefont{Kasprzak et~al.}(2006)\citenamefont{Kasprzak, Richard,
  Kundermann, Baas, Jeambrun, Keeling, Marchetti, Szymanska, Andr\'e, Staehli
  et~al.}}]{kasprzak06a}
\bibinfo{author}{\bibfnamefont{J.}~\bibnamefont{Kasprzak}},
  \bibinfo{author}{\bibfnamefont{M.}~\bibnamefont{Richard}},
  \bibinfo{author}{\bibfnamefont{S.}~\bibnamefont{Kundermann}},
  \bibinfo{author}{\bibfnamefont{A.}~\bibnamefont{Baas}},
  \bibinfo{author}{\bibfnamefont{P.}~\bibnamefont{Jeambrun}},
  \bibinfo{author}{\bibfnamefont{J.~M.~J.} \bibnamefont{Keeling}},
  \bibinfo{author}{\bibfnamefont{F.~M.} \bibnamefont{Marchetti}},
  \bibinfo{author}{\bibfnamefont{M.~H.} \bibnamefont{Szymanska}},
  \bibinfo{author}{\bibfnamefont{R.}~\bibnamefont{Andr\'e}},
  \bibinfo{author}{\bibfnamefont{J.~L.} \bibnamefont{Staehli}},
  \bibnamefont{et~al.}, \bibinfo{journal}{Nature}
  \textbf{\bibinfo{volume}{443}}, \bibinfo{pages}{409} (\bibinfo{year}{2006}).

\bibitem[{\citenamefont{Amo et~al.}(2009{\natexlab{a}})\citenamefont{Amo,
  Lefr\`ere, Pigeon, Adrados, Ciuti, Carusotto, Houdr\'e, Giacobino, and
  Bramati}}]{amo09b}
\bibinfo{author}{\bibfnamefont{A.}~\bibnamefont{Amo}},
  \bibinfo{author}{\bibfnamefont{J.}~\bibnamefont{Lefr\`ere}},
  \bibinfo{author}{\bibfnamefont{S.}~\bibnamefont{Pigeon}},
  \bibinfo{author}{\bibfnamefont{C.}~\bibnamefont{Adrados}},
  \bibinfo{author}{\bibfnamefont{C.}~\bibnamefont{Ciuti}},
  \bibinfo{author}{\bibfnamefont{I.}~\bibnamefont{Carusotto}},
  \bibinfo{author}{\bibfnamefont{R.}~\bibnamefont{Houdr\'e}},
  \bibinfo{author}{\bibfnamefont{E.}~\bibnamefont{Giacobino}},
  \bibnamefont{and} \bibinfo{author}{\bibfnamefont{A.}~\bibnamefont{Bramati}},
  \bibinfo{journal}{Nature Phys.} \textbf{\bibinfo{volume}{5}},
  \bibinfo{pages}{805} (\bibinfo{year}{2009}{\natexlab{a}}).

\bibitem[{\citenamefont{Amo et~al.}(2009{\natexlab{b}})\citenamefont{Amo,
  Sanvitto, Laussy, Ballarini, del Valle, Martin, Lema\^itre, Bloch,
  Krizhanovskii, Skolnick et~al.}}]{amo09a}
\bibinfo{author}{\bibfnamefont{A.}~\bibnamefont{Amo}},
  \bibinfo{author}{\bibfnamefont{D.}~\bibnamefont{Sanvitto}},
  \bibinfo{author}{\bibfnamefont{F.~P.} \bibnamefont{Laussy}},
  \bibinfo{author}{\bibfnamefont{D.}~\bibnamefont{Ballarini}},
  \bibinfo{author}{\bibfnamefont{E.}~\bibnamefont{del Valle}},
  \bibinfo{author}{\bibfnamefont{M.~D.} \bibnamefont{Martin}},
  \bibinfo{author}{\bibfnamefont{A.}~\bibnamefont{Lema\^itre}},
  \bibinfo{author}{\bibfnamefont{J.}~\bibnamefont{Bloch}},
  \bibinfo{author}{\bibfnamefont{D.~N.} \bibnamefont{Krizhanovskii}},
  \bibinfo{author}{\bibfnamefont{M.~S.} \bibnamefont{Skolnick}},
  \bibnamefont{et~al.}, \bibinfo{journal}{Nature}
  \textbf{\bibinfo{volume}{457}}, \bibinfo{pages}{291}
  (\bibinfo{year}{2009}{\natexlab{b}}).

\bibitem[{\citenamefont{Baumberg et~al.}(2008)\citenamefont{Baumberg, Kavokin,
  Christopoulos, Grundy, Butt\'e, Christmann, Solnyshkov, Malpuech, von
  H\"ogersthal, Feltin et~al.}}]{baumberg08a}
\bibinfo{author}{\bibfnamefont{J.~J.} \bibnamefont{Baumberg}},
  \bibinfo{author}{\bibfnamefont{A.~V.} \bibnamefont{Kavokin}},
  \bibinfo{author}{\bibfnamefont{S.}~\bibnamefont{Christopoulos}},
  \bibinfo{author}{\bibfnamefont{A.~J.~D.} \bibnamefont{Grundy}},
  \bibinfo{author}{\bibfnamefont{R.}~\bibnamefont{Butt\'e}},
  \bibinfo{author}{\bibfnamefont{G.}~\bibnamefont{Christmann}},
  \bibinfo{author}{\bibfnamefont{D.~D.} \bibnamefont{Solnyshkov}},
  \bibinfo{author}{\bibfnamefont{G.}~\bibnamefont{Malpuech}},
  \bibinfo{author}{\bibfnamefont{G.~B.~H.} \bibnamefont{von H\"ogersthal}},
  \bibinfo{author}{\bibfnamefont{E.}~\bibnamefont{Feltin}},
  \bibnamefont{et~al.}, \bibinfo{journal}{Phys. Rev. Lett.}
  \textbf{\bibinfo{volume}{101}}, \bibinfo{pages}{136409}
  (\bibinfo{year}{2008}).

\bibitem[{\citenamefont{Kavokin et~al.}(2003)\citenamefont{Kavokin, Malpuech,
  and Laussy}}]{kavokin03a}
\bibinfo{author}{\bibfnamefont{A.}~\bibnamefont{Kavokin}},
  \bibinfo{author}{\bibfnamefont{G.}~\bibnamefont{Malpuech}}, \bibnamefont{and}
  \bibinfo{author}{\bibfnamefont{F.~P.} \bibnamefont{Laussy}},
  \bibinfo{journal}{Phys. Lett. A} \textbf{\bibinfo{volume}{306}},
  \bibinfo{pages}{187} (\bibinfo{year}{2003}).

\bibitem[{\citenamefont{Lagoudakis et~al.}(2003)\citenamefont{Lagoudakis,
  Martin, Baumberg, Qarry, Cohen, and Pfeiffer}}]{lagoudakis03a}
\bibinfo{author}{\bibfnamefont{P.~G.} \bibnamefont{Lagoudakis}},
  \bibinfo{author}{\bibfnamefont{M.~D.} \bibnamefont{Martin}},
  \bibinfo{author}{\bibfnamefont{J.~J.} \bibnamefont{Baumberg}},
  \bibinfo{author}{\bibfnamefont{A.}~\bibnamefont{Qarry}},
  \bibinfo{author}{\bibfnamefont{E.}~\bibnamefont{Cohen}}, \bibnamefont{and}
  \bibinfo{author}{\bibfnamefont{L.~N.} \bibnamefont{Pfeiffer}},
  \bibinfo{journal}{Phys. Rev. Lett.} \textbf{\bibinfo{volume}{90}},
  \bibinfo{pages}{206401} (\bibinfo{year}{2003}).

\bibitem[{\citenamefont{Bardeen et~al.}(1957)\citenamefont{Bardeen, Cooper, and
  Schrieffer}}]{bardeen57b}
\bibinfo{author}{\bibfnamefont{J.}~\bibnamefont{Bardeen}},
  \bibinfo{author}{\bibfnamefont{L.~N.} \bibnamefont{Cooper}},
  \bibnamefont{and} \bibinfo{author}{\bibfnamefont{J.~R.}
  \bibnamefont{Schrieffer}}, \bibinfo{journal}{Phys. Rev.}
  \textbf{\bibinfo{volume}{106}}, \bibinfo{pages}{162} (\bibinfo{year}{1957}).

\bibitem[{\citenamefont{Bardeen}(1951)}]{bardeen51a}
\bibinfo{author}{\bibfnamefont{J.}~\bibnamefont{Bardeen}},
  \bibinfo{journal}{Rev. Mod. Phys.} \textbf{\bibinfo{volume}{23}},
  \bibinfo{pages}{261} (\bibinfo{year}{1951}).

\bibitem[{\citenamefont{Cooper}(1956)}]{cooper56a}
\bibinfo{author}{\bibfnamefont{L.~N.} \bibnamefont{Cooper}},
  \bibinfo{journal}{Phys. Rev.} \textbf{\bibinfo{volume}{104}},
  \bibinfo{pages}{1189} (\bibinfo{year}{1956}).

\bibitem[{\citenamefont{Giustino et~al.}(2008)\citenamefont{Giustino, Cohen,
  and Louie}}]{giustino08a}
\bibinfo{author}{\bibfnamefont{F.}~\bibnamefont{Giustino}},
  \bibinfo{author}{\bibfnamefont{M.~L.} \bibnamefont{Cohen}}, \bibnamefont{and}
  \bibinfo{author}{\bibfnamefont{S.~G.} \bibnamefont{Louie}},
  \bibinfo{journal}{Nature} \textbf{\bibinfo{volume}{452}},
  \bibinfo{pages}{975} (\bibinfo{year}{2008}).

\bibitem[{\citenamefont{Little}(1964)}]{little64a}
\bibinfo{author}{\bibfnamefont{W.~A.} \bibnamefont{Little}},
  \bibinfo{journal}{Phys. Rev.} \textbf{\bibinfo{volume}{134}},
  \bibinfo{pages}{A1416} (\bibinfo{year}{1964}).

\bibitem[{\citenamefont{Ginzburg}(2009)}]{ginzburg_book09a}
\bibinfo{author}{\bibfnamefont{V.~L.} \bibnamefont{Ginzburg}},
  \emph{\bibinfo{title}{On Superconductivity and Superfluidity}}
  (\bibinfo{publisher}{Springer}, \bibinfo{year}{2009}).

\bibitem[{\citenamefont{Allender et~al.}(1973)\citenamefont{Allender, Bray, and
  Bardeen}}]{allender73a}
\bibinfo{author}{\bibfnamefont{D.}~\bibnamefont{Allender}},
  \bibinfo{author}{\bibfnamefont{J.}~\bibnamefont{Bray}}, \bibnamefont{and}
  \bibinfo{author}{\bibfnamefont{J.}~\bibnamefont{Bardeen}},
  \bibinfo{journal}{Phys. Rev. B} \textbf{\bibinfo{volume}{7}},
  \bibinfo{pages}{1020} (\bibinfo{year}{1973}).

\bibitem[{\citenamefont{Aruta et~al.}(2008)\citenamefont{Aruta, Ghiringhelli,
  Dallera, Fracassi, Medaglia, Tebano, Brookes, Braicovich, and
  Balestrino}}]{aruta08a}
\bibinfo{author}{\bibfnamefont{C.}~\bibnamefont{Aruta}},
  \bibinfo{author}{\bibfnamefont{G.}~\bibnamefont{Ghiringhelli}},
  \bibinfo{author}{\bibfnamefont{C.}~\bibnamefont{Dallera}},
  \bibinfo{author}{\bibfnamefont{F.}~\bibnamefont{Fracassi}},
  \bibinfo{author}{\bibfnamefont{P.~G.} \bibnamefont{Medaglia}},
  \bibinfo{author}{\bibfnamefont{A.}~\bibnamefont{Tebano}},
  \bibinfo{author}{\bibfnamefont{N.~B.} \bibnamefont{Brookes}},
  \bibinfo{author}{\bibfnamefont{L.}~\bibnamefont{Braicovich}},
  \bibnamefont{and}
  \bibinfo{author}{\bibfnamefont{G.}~\bibnamefont{Balestrino}},
  \bibinfo{journal}{Phys. Rev. B} \textbf{\bibinfo{volume}{78}},
  \bibinfo{pages}{205120} (\bibinfo{year}{2008}).

\bibitem[{\citenamefont{Gozar et~al.}(2008)\citenamefont{Gozar, Logvenov,
  Kourkoutis, Bollinger, Giannuzzi, Muller, and Bozovic}}]{gozar08a}
\bibinfo{author}{\bibfnamefont{A.}~\bibnamefont{Gozar}},
  \bibinfo{author}{\bibfnamefont{G.}~\bibnamefont{Logvenov}},
  \bibinfo{author}{\bibfnamefont{L.~F.} \bibnamefont{Kourkoutis}},
  \bibinfo{author}{\bibfnamefont{A.~T.} \bibnamefont{Bollinger}},
  \bibinfo{author}{\bibfnamefont{L.~A.} \bibnamefont{Giannuzzi}},
  \bibinfo{author}{\bibfnamefont{D.~A.} \bibnamefont{Muller}},
  \bibnamefont{and} \bibinfo{author}{\bibfnamefont{I.}~\bibnamefont{Bozovic}},
  \bibinfo{journal}{Nature} \textbf{\bibinfo{volume}{455}},
  \bibinfo{pages}{782} (\bibinfo{year}{2008}).

\bibitem[{\citenamefont{Ginsburg and Kirzhnits}(1977)}]{ps_ginsburg77}
\bibinfo{editor}{\bibfnamefont{V.}~\bibnamefont{Ginsburg}} \bibnamefont{and}
  \bibinfo{editor}{\bibfnamefont{K.}~\bibnamefont{Kirzhnits}}, eds.,
  \emph{\bibinfo{title}{High temperature superconductivity}}
  (\bibinfo{publisher}{Pergamon}, \bibinfo{year}{1977}).

\bibitem[{\citenamefont{Tassone and Yamamoto}(1999)}]{tassone99a}
\bibinfo{author}{\bibfnamefont{F.}~\bibnamefont{Tassone}} \bibnamefont{and}
  \bibinfo{author}{\bibfnamefont{Y.}~\bibnamefont{Yamamoto}},
  \bibinfo{journal}{Phys. Rev. B} \textbf{\bibinfo{volume}{59}},
  \bibinfo{pages}{10830} (\bibinfo{year}{1999}).

\bibitem[{\citenamefont{Ciuti et~al.}(1998)\citenamefont{Ciuti, Savona,
  Piermarocchi, Quattropani, and Schwendimann}}]{ciuti98a}
\bibinfo{author}{\bibfnamefont{C.}~\bibnamefont{Ciuti}},
  \bibinfo{author}{\bibfnamefont{V.}~\bibnamefont{Savona}},
  \bibinfo{author}{\bibfnamefont{C.}~\bibnamefont{Piermarocchi}},
  \bibinfo{author}{\bibfnamefont{A.}~\bibnamefont{Quattropani}},
  \bibnamefont{and}
  \bibinfo{author}{\bibfnamefont{P.}~\bibnamefont{Schwendimann}},
  \bibinfo{journal}{Phys. Rev. B} \textbf{\bibinfo{volume}{58}},
  \bibinfo{pages}{7926} (\bibinfo{year}{1998}).

\bibitem[{\citenamefont{Haug and Koch}(1990)}]{haug_book90a}
\bibinfo{author}{\bibfnamefont{H.}~\bibnamefont{Haug}} \bibnamefont{and}
  \bibinfo{author}{\bibfnamefont{S.~W.} \bibnamefont{Koch}},
  \emph{\bibinfo{title}{Quantum theory of the optical and electronic properties
  of semiconductors}} (\bibinfo{publisher}{World Scientific},
  \bibinfo{year}{1990}).

\bibitem[{\citenamefont{Bijlsma et~al.}(2000)\citenamefont{Bijlsma, Heringa,
  and Stoof}}]{bijlsma00a}
\bibinfo{author}{\bibfnamefont{M.~J.} \bibnamefont{Bijlsma}},
  \bibinfo{author}{\bibfnamefont{B.~A.} \bibnamefont{Heringa}},
  \bibnamefont{and} \bibinfo{author}{\bibfnamefont{H.~T.~C.}
  \bibnamefont{Stoof}}, \bibinfo{journal}{Phys. Rev. A}
  \textbf{\bibinfo{volume}{61}}, \bibinfo{pages}{053601}
  (\bibinfo{year}{2000}).

\bibitem[{\citenamefont{Christmann et~al.}(2008)\citenamefont{Christmann,
  Butt\'e, Feltin, Carlin, and Grandjean}}]{christmann08a}
\bibinfo{author}{\bibfnamefont{G.}~\bibnamefont{Christmann}},
  \bibinfo{author}{\bibfnamefont{R.}~\bibnamefont{Butt\'e}},
  \bibinfo{author}{\bibfnamefont{E.}~\bibnamefont{Feltin}},
  \bibinfo{author}{\bibfnamefont{J.-F.} \bibnamefont{Carlin}},
  \bibnamefont{and}
  \bibinfo{author}{\bibfnamefont{N.}~\bibnamefont{Grandjean}},
  \bibinfo{journal}{Appl. Phys. Lett.} \textbf{\bibinfo{volume}{93}},
  \bibinfo{pages}{051102} (\bibinfo{year}{2008}).

\end{thebibliography}

\end{document}